\documentclass[final]{cvpr}

\usepackage{times}
\usepackage{epsfig}
\usepackage{graphicx}
\usepackage{amsmath}
\usepackage{amssymb}
\usepackage{booktabs}
\usepackage{multirow}

\usepackage[pagebackref=true,breaklinks=true,colorlinks,bookmarks=false]{hyperref}

\def\cvprPaperID{****} 
\def\confYear{CVPR 2021}

\begin{document}

\title{NTIRE 2021 Challenge on Perceptual Image Quality Assessment}

\author{
Jinjin Gu \and Haoming Cai \and Chao Dong \and Jimmy S. Ren \and Yu Qiao  \and Shuhang Gu \and Radu Timofte \and
Manri Cheon \and Sungjun Yoon \and Byungyeon Kang \and Junwoo Lee \and
Qing Zhang \and Haiyang Guo \and Yi Bin \and Yuqing Hou \and Hengliang Luo \and 
Jingyu Guo \and Zirui Wang \and Hai Wang \and Wenming Yang \and
Qingyan Bai \and Shuwei Shi \and Weihao Xia \and Mingdeng Cao \and Jiahao Wang \and Yifan Chen \and Yujiu Yang \and
Yang Li \and Tao Zhang \and Longtao Feng \and Yiting Liao \and Junlin Li \and
William Thong \and Jose Costa Pereira \and Ales Leonardis \and Steven McDonagh \and 
Kele Xu \and Lehan Yang \and Hengxing Cai \and Pengfei Sun \and
Seyed Mehdi Ayyoubzadeh \and Ali Royat \and 
Sid Ahmed Fezza \and Dounia Hammou \and Wassim Hamidouche \and
Sewoong Ahn \and Kwangjin Yoon \and
Koki Tsubota \and Hiroaki Akutsu \and Kiyoharu Aizawa
}

\maketitle

\begin{abstract}
This paper reports on the NTIRE 2021 challenge on perceptual image quality assessment (IQA), held in conjunction with the New Trends in Image Restoration and Enhancement workshop (NTIRE) workshop at CVPR 2021.
As a new type of image processing technology, perceptual image processing algorithms based on Generative Adversarial Networks (GAN) have produced images with more realistic textures.
These output images have completely different characteristics from traditional distortions, thus pose a new challenge for IQA methods to evaluate their visual quality.
In comparison with previous IQA challenges, the training and testing datasets in this challenge include the outputs of perceptual image processing algorithms and the corresponding subjective scores. Thus they can be used to develop and evaluate IQA methods on GAN-based distortions.
The challenge has 270 registered participants in total. In the final testing stage, 13 participating teams submitted their models and fact sheets.
Almost all of them have achieved much better results than existing IQA methods, while the winning method can demonstrate state-of-the-art performance.
\end{abstract}
{\let\thefootnote\relax\footnotetext{%
\hspace{-5mm}$^*$Jinjin Gu (\texttt{jinjin.gu@sydney.edu.au}), Haoming Cai, Chao Dong, Jimmy Ren, Yu Qiao, Shuhang Gu and Radu Timofte are the NTIRE 2021 challenge organizers. The other authors participated in the challenge.\\
Appendix.\ref{sec:apd:team} contains the authors' team names and affiliations.\\The NTIRE website: \url{https://data.vision.ee.ethz.ch/cvl/ntire21/}
}}

\begin{figure}
    \centering
    \includegraphics[width=\linewidth]{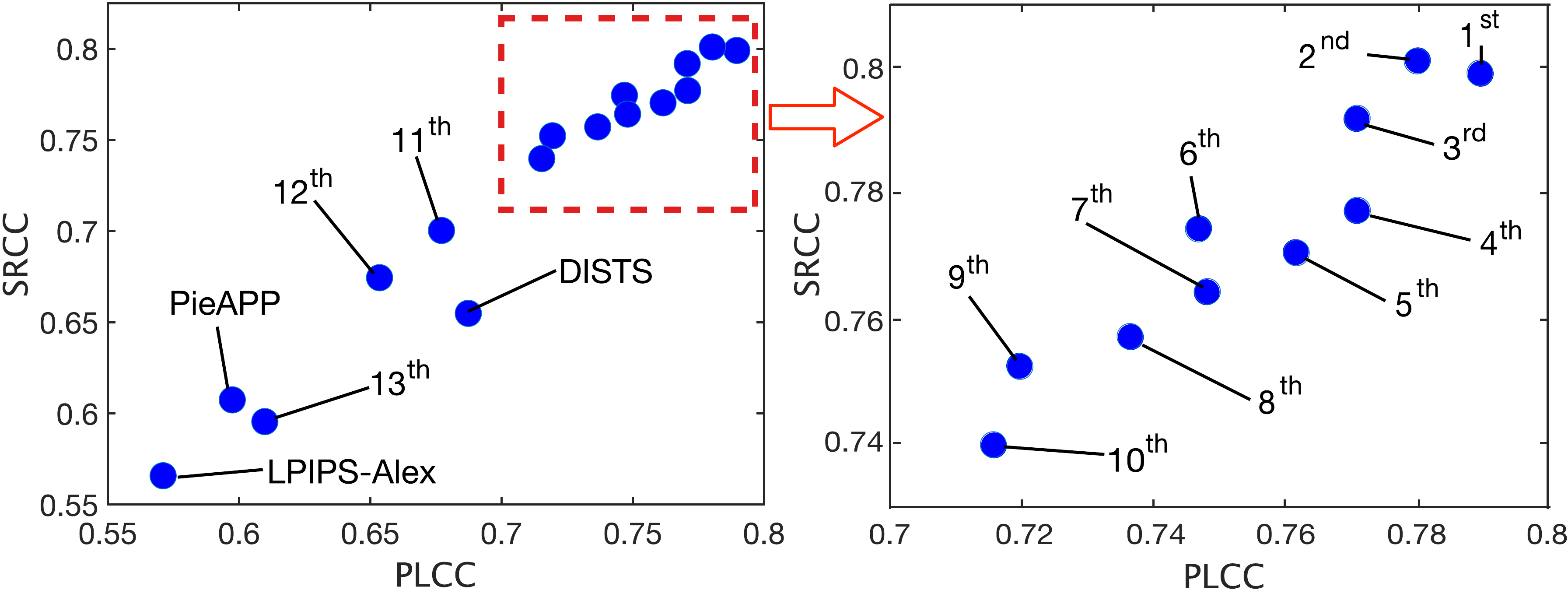}
    \caption{Quantitative comparison of IQA methods in the challenge. The right figure is the zoom-in view. SRCC represents Spearman rank order correlation coefficient and PLCC represents Pearson linear correlation coefficient. Higher coefficient matches perceptual score better. The top methods demonstrate the state-of-the-art performance.}
    \label{fig:teaser}
\end{figure}

\section{Introduction}
Image quality assessment (IQA) aims at using computational models to measure the perceptual quality of images, which are degraded during acquisition, compression, reproduction and post-processing operations.
As the ``evaluation mechanism'', IQA plays a critical role in most image processing tasks, such as image super-resolution, denoising, compression and enhancement.
Although it is easy for human beings to distinguish perceptually better images, it has been proved to be difficult for algorithms \cite{tid2013,pipal}.
Especially, on the basis of Generative Adversarial Networks (GANs) \cite{goodfellow2014generative}, perceptual image processing algorithms (or perceptual-oriented algorithms) \cite{johnson2016perceptual,srgan2017,wang2018esrgan,zhang2019ranksrgan} have posed a great challenge for IQA methods, as they bring completely new characteristics to the output images \cite{pipal}.
It has been noticed that the contradiction between the quantitative evaluation results and the real perceptual quality is increasing \cite{blau20182018,blau2018perception,pipal}.
This will also affect the development of image processing algorithms, if the IQA methods cannot objectively compare their perceptual quality \cite{blau2018perception,pipal}.
Therefore, new IQA methods need to be proposed accordingly, to adapt new image processing algorithms.

The NTIRE 2021 challenge takes a step forward in benchmarking perceptual IQA -- the task of predicting the perceptual quality of an image obtained by perceptual-oriented algorithms.
We employ a new dataset called PIPAL \cite{pipal} as our training set, which contains 200 reference images, 29k distorted images and 1.13M human judgements.
Especially, this dataset includes the results of perceptual-oriented algorithms, which are missing in previous datasets.
We also collect an extended dataset of PIPAL for validation and testing. This dataset contains 3,300 distorted images for 50 reference images, and all of them are the outputs of perceptual-oriented algorithms.
We collect 753k human judgements to assign subjective scores for the extended images, ensuring the objectivity of the testing data.

The challenge has 270 registered participants in total. Among them, 13 participating teams have submitted their final solutions and fact sheets.
They introduce new technologies in network architectures, loss functions, ensemble methods, data augmentation methods, and \etc.
The performance overview of these solutions is shown in \figurename~\ref{fig:teaser}.

This challenge is one of the NTIRE 2021 associated challenges: nonhomogeneous dehazing~\cite{ancuti2021ntire}, defocus deblurring using dual-pixel~\cite{abuolaim2021ntire}, depth guided image relighting~\cite{elhelou2021ntire}, image deblurring~\cite{nah2021ntire}, multi-modal aerial view imagery classification~\cite{liu2021ntire}, learning the super-resolution space~\cite{lugmayr2021ntire}, quality enhancement of heavily compressed videos~\cite{yang2021ntire}, video super-resolution~\cite{son2021ntire}, perceptual image quality assessment~\cite{gu2021ntire}, burst super-resolution~\cite{bhat2021ntire}, high dynamic range~\cite{perez2021ntire}.

\section{Related Work}
\paragraph{Image quality assessment (IQA).}
According to different usage scenarios, IQA methods can be divided in to full-reference methods (FR-IQA) and no-reference methods (NR-IQA).
FR-IQA methods measure the perceptual similarity between two images, and have been widely used in the evaluation of image/video coding, restoration and communication quality.
Beyond the most widely-used PSNR, FR-IQA methods follow a long line of works that can trace back to SSIM \cite{ssim}, which first introduces structural information in measuring image similarity.
SSIM opened a precedent for the evaluation of image structure or features.
After that, various FR-IQA methods have been proposed to bridge the gap between results of IQA methods and human judgements \cite{ms-ssim,sr-sim,fsim,ifc,vsi}.
Similar to other computer vision problems, advanced data-driven methods have also motivated the investigation of applications of IQA.
Zhang \etal \cite{zhang2018unreasonable} propose to use pre-trained deep networks to calculate the perceptual similarity and achieve good results.
A contemporaneous work \cite{prashnani2018pieapp} also propose to train deep IQA network using a pairwise-learning framework to predict the preference of one distorted image over the other.
In addition to the above FR-IQA methods, NR-IQA methods are proposed to assess image quality without a reference image.
Some popular NR-IQA methods include NIQE \cite{niqe}, \cite{ma2017learning}, BRISQUE \cite{brisque}, and PI \cite{blau2018perception}.

\begin{figure}
    \centering
    \includegraphics[width=\linewidth]{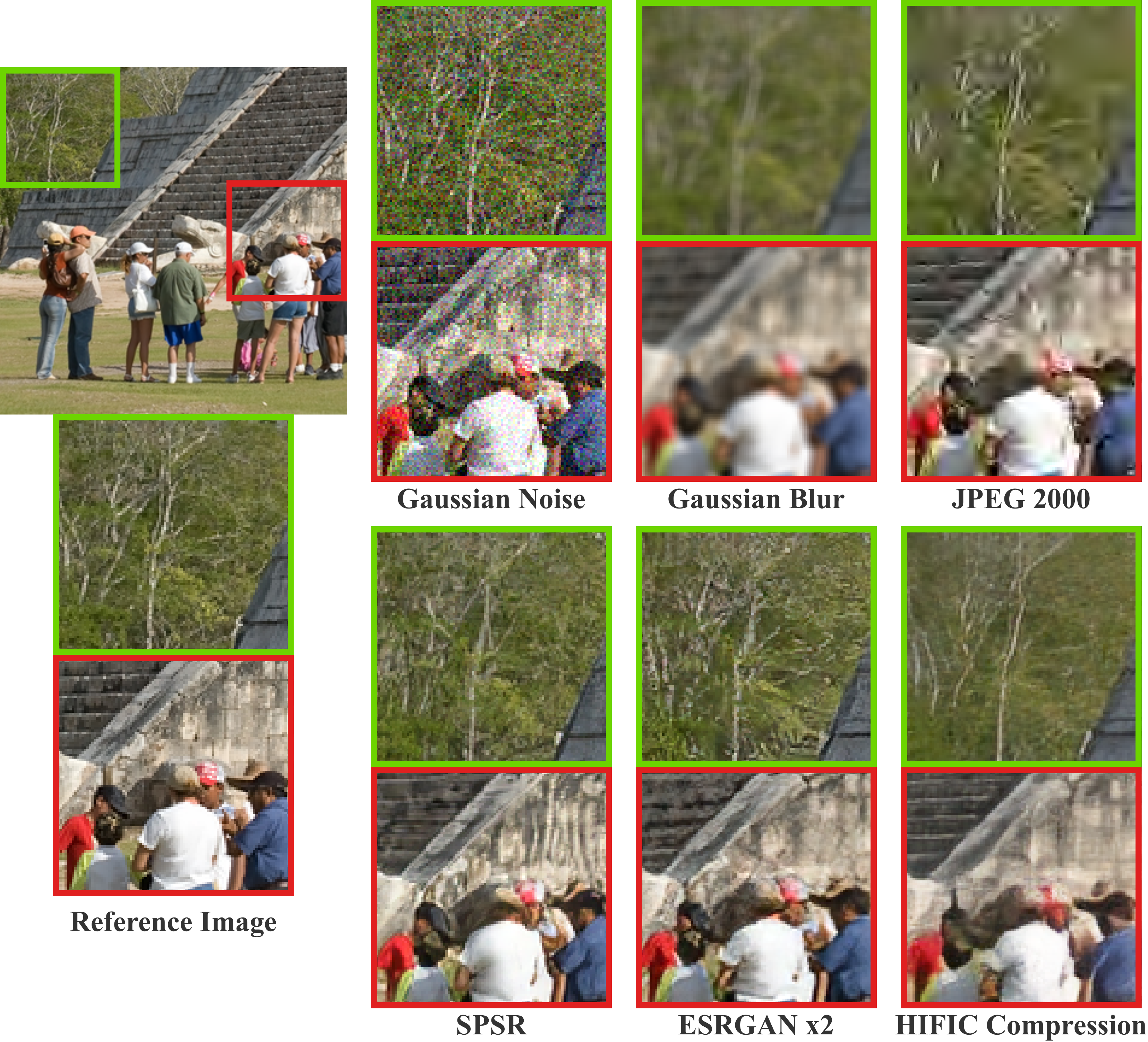}
    \caption{The difference between the traditional distortions (the first row) and the outputs of perceptual-oriented algorithms (the second row).}
    \label{fig:gan_demo}
\end{figure}

\begin{table*}[t]
    \centering
    \caption{Quantitative results for the NTIRE 2021 Perceptual IQA challenge.}
    \label{tab:main_results}
    \vspace{2mm}
    \resizebox{1.0\linewidth}{!}{
    \begin{tabular}{c|l|l||c|cc|cc|cc}
    \toprule
        \multirow{2}{*}{Rank} & \multirow{2}{*}{Team Name} & \multirow{2}{*}{Author/Method} & \multirow{2}{*}{Main Score} & \multicolumn{2}{c}{PIPAL} & \multicolumn{2}{c}{TID2013} & \multicolumn{2}{c}{LIVE} \\
        & & & & SRCC & PLCC & SRCC & PLCC & SRCC & PLCC\\
        \midrule
        1 & LIPT              & mrc         & 1.5885 & 0.7990 & 0.7896  & 0.8040 & 0.8440 & 0.9170 & 0.8970 \\
        2 & MT-GTD            & binyi       & 1.5811 & 0.8009 & 0.7803  & 0.7815 & 0.8265 & 0.9191 & 0.9132 \\
        3 & The Amaurotia     & gjy19       & 1.5625 & 0.7918 & 0.7707  & 0.7293 & 0.7931 & 0.9084 & 0.9074 \\
        4 & THUIIGROUP1919    & bqy2020     & 1.5480 & 0.7770 & 0.7709  & 0.7465 & 0.7960 & 0.9051 & 0.8777 \\
        5 & Yahaha!           & sherlocky   & 1.5317 & 0.7703 & 0.7615  & —     & —     & —     & —     \\
        6 & Huawei Noah's Ark & wth         & 1.5212 & 0.7744 & 0.7468  & —     & —     & —     & —     \\
        7 & debut\_kele       & debut\_kele & 1.5121 & 0.7641 & 0.7480  & —     & —     & —     & —     \\
        8 & zhangtaotao       & zhangtaotao & 1.4936 & 0.7571 & 0.7366  & —     & —     & —     & —     \\
        9 & MACS              & alir        & 1.4717 & 0.7522 & 0.7194  & 0.7300 & 0.7800 & 0.9200 & 0.9400 \\
        10& orboai            & orboai      & 1.4549 & 0.7397 & 0.7153  & —     & —     & —     & —     \\
        11& LION Team         & sfezza      & 1.3774 & 0.7003 & 0.6771  & —     & —     & —     & —     \\
        12& SI analytics      & Ahn         & 1.3280 & 0.6744 & 0.6535  & 0.5371 & 0.6264 & 0.9263 & 0.9211 \\
        13& tsubota           & tsubota     & 1.2053 & 0.5955 & 0.6098  & —     & —     & —     & —     \\
        \midrule
        \multirow{8}{*}{} & \multirow{8}{*}{Baselines} 
        & LPIPS-VGG  & 1.2277 & 0.5947 & 0.6330  & 0.6695 & 0.7490 & 0.9433 & 0.9431 \\
        & & LPIPS-Alex & 1.1368 & 0.5658 & 0.5711  & 0.7444 & 0.7634 & 0.9211 & 0.9172 \\
        & & PieAPP     & 1.2048 & 0.6074 & 0.5974  & 0.8478 & 0.8064 & 0.9182 & 0.9102 \\
        & & DISTS      & 1.3422 & 0.6548 & 0.6873  & 0.8184 & 0.8463 & 0.9468 & 0.9440 \\
        & & SWD        & 1.2585 & 0.6243 & 0.6342  & 0.7895 & 0.8219 & 0.8832 & 0.8731 \\
        & & FSIM       & 1.0748 & 0.5038 & 0.5709  & 0.8015 & 0.8560 & 0.9634 & 0.9491 \\
        & & SSIM       & 0.7549 & 0.3614 & 0.3936  & 0.7414 & 0.7894 & 0.9479 & 0.9397 \\
        & & PSNR       & 0.5263 & 0.2493 & 0.2769  & 0.6395 & 0.6541 & 0.8756 & 0.8686 \\
        \bottomrule
    \end{tabular}
    }
\end{table*}

\paragraph{Perceptual-oriented and GAN-based distortion.}
In the past years, photo-realistic image generation has been evolving rapidly \cite{srgan2017,wang2018esrgan,sftgan2018,zhang2019ranksrgan}, benefiting from the invention of perceptual-oriented loss function \cite{johnson2016perceptual,wang2018esrgan} and GANs \cite{goodfellow2014generative}.
On the one hand, this kind of perceptual image restoration algorithm greatly improves the perceptual effect of the output image.
On the other hand, it brings completely new characteristics to the output images.
In general, these methods often fabricate seemingly realistic yet fake details and textures.
They do not quite match the quality of detail loss, as they usually contain texture-like noise, or the quality of noise, the noise is similar to the ground truth in appearance but is not accurate.
An example of perceptual distortions is shown in \figurename~\ref{fig:gan_demo}.
The quality evaluation of such images has been proved challenging for IQA methods \cite{pipal}.
In order to evaluate and improve the performance of the IQA method against such perceptual distortions, Gu \etal \cite{pipal} contribute a new IQA dataset called Perceptual Image Processing ALgorithms dataset (PIPAL), including the results of Perceptual-oriented image processing algorithms, which are missing in previous datasets. 
Recently, Gu \etal \cite{gu2020image} propose to improve the IQA performance on these perceptual distortions by explicitly considering the spatial misalignment using anti-aliasing pooling layers and spatially robust comparison operations in the IQA network.

\begin{figure*}
    \centering
    \includegraphics[width=\linewidth]{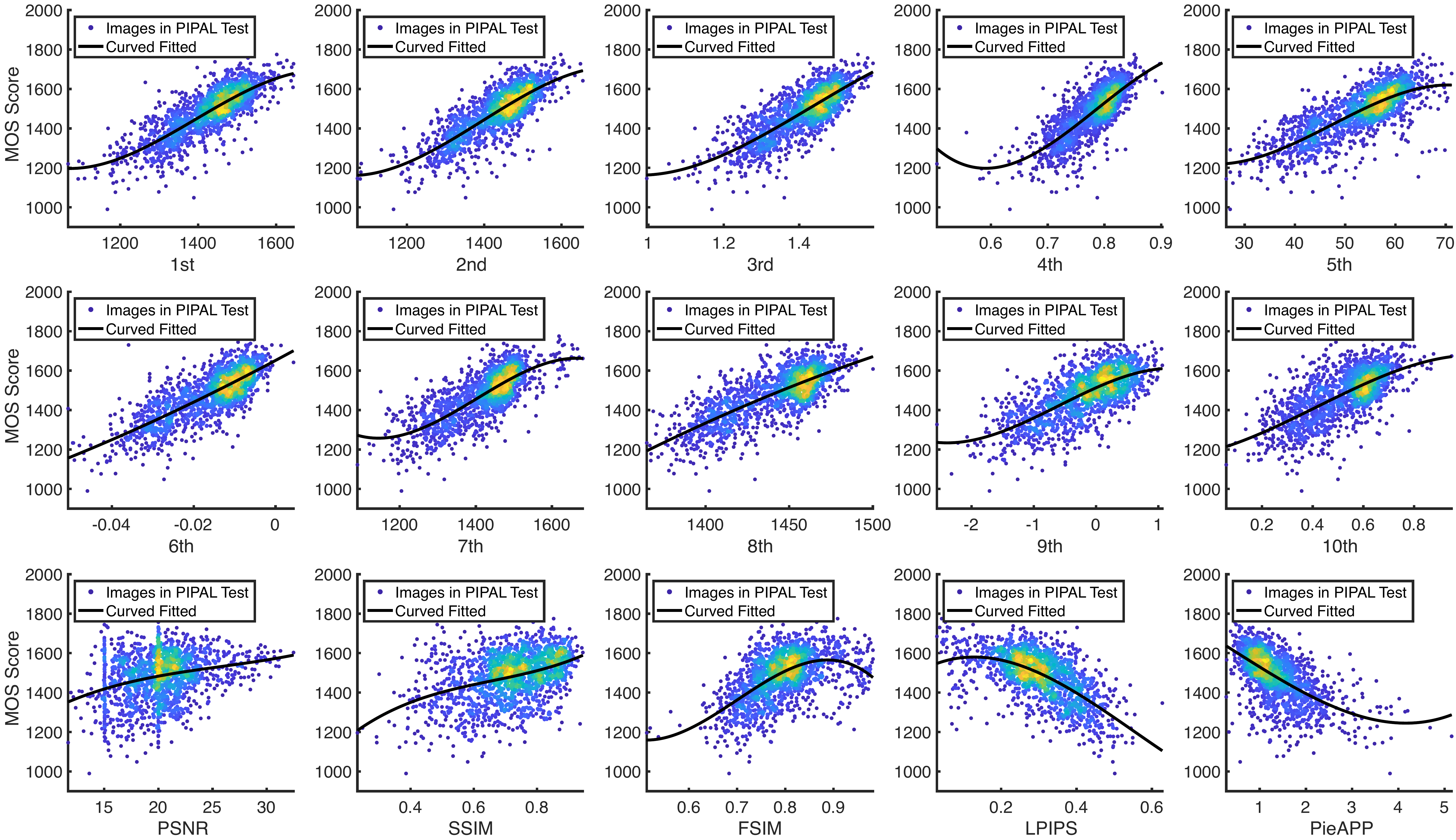}
    \caption{Scatter plots of the objective scores vs. the MOS scores.}
    \label{fig:scatter_plots}
\end{figure*}

\section{The NTIRE Challenge on Perceptual IQA}
We host the NTIRE 2021 Perceptual Image Quality Assessment Challenge and the objective are three-fold:
(1) to push developing state-of-the-art perceptual image quality assessment algorithms to deal with the novel GAN-based distortion types;
(2) to compare different solutions and gain new insights;
and (3) to promote a novel large perceptual IQA dataset (PIPAL \cite{pipal,gu2020image}).
Details about the challenge are as follows:

\paragraph{Task.}
The task of this challenge is to obtain an image quality assessment method capable to produce high-quality perceptual similarity results between the given distorted images and the corresponding reference images with the best correlation to the reference ground truth MOS score.
Note that we do not restrict the participants to develop the full-reference IQA methods, and the blind IQA methods are also welcomed.

\paragraph{Dataset.}
We employ a subset of the PIPAL dataset as the training set and an extended version of the PIPAL dataset as the validation and the testing set.
The PIPAL dataset includes both traditional distortion types, image restoration results, compression results, and novel GAN-based image processing outputs.
More than 1.13 million human judgements are collected to assign mean opinion scores (MOS) for PIPAL images using the Elo rating system \cite{elo1978rating}.
The original PIPAL dataset includes 250 high-quality diverse reference images, each has 116 different distorted images.
We use 200 of the 250 reference images and their distorted images as the training set (in total $200\times116$ distorted images).
All training images and the MOS scores are publicly available.

We collect an extended version of the PIPAL dataset as the validation and testing set.
We use the rest 50 reference images and collect 66 additional distorted images for each of them.
The newly collected distortion types are all outputs of GAN-based image restoration algorithms or GAN-based compression algorithms, in total 3300 additional images are collected.
Thanks to the expandability of the Elo rating system used by the PIPAL dataset, we can assign MOS scores to new images with additional pairwise judgements without collecting from scratch.
At last, 753k human judgements are involved in preparing the validation and testing set.
The validation set contains 25 reference images and 40 distorted images for each of them.
The testing set contains the rest 25 reference images and all the 66 distorted images for each reference image.
Note that for the participants, the training set and the validation/testing set contain completely different reference and distorted images, which ensures the objectivity of the final results.

\begin{figure*}
    \centering
    \includegraphics[width=\linewidth]{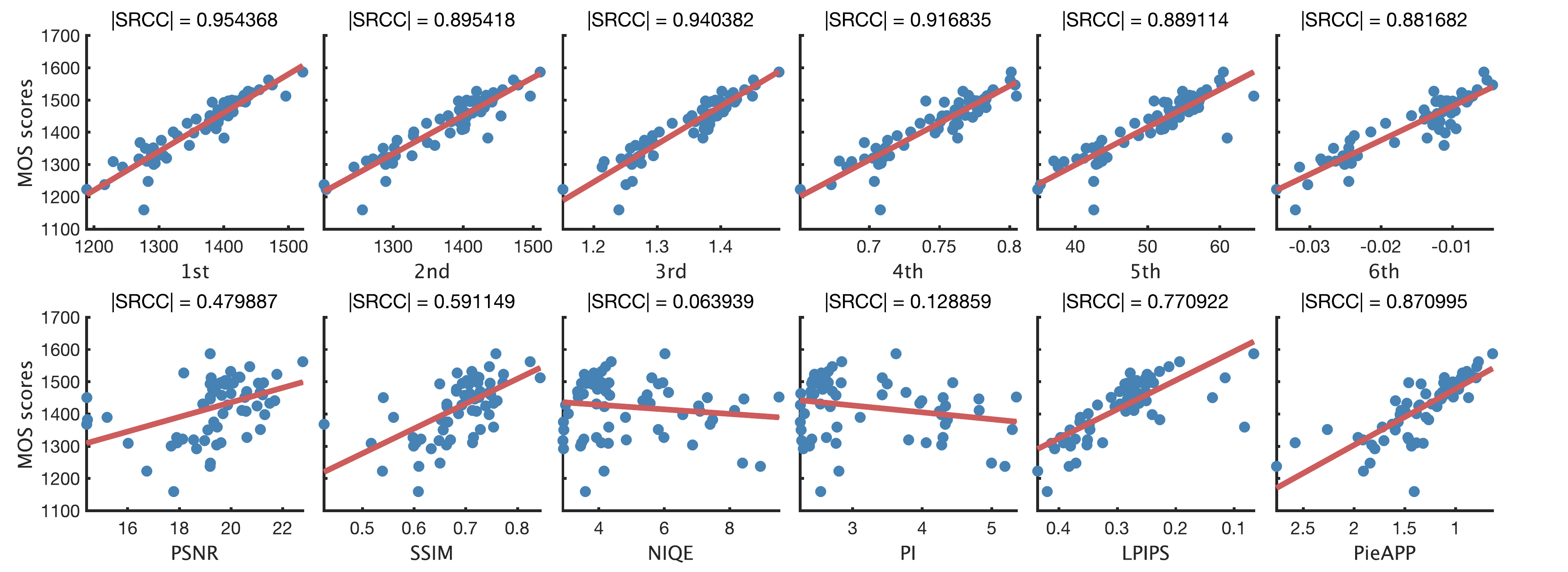}
    \caption{Analysis of IQA methods in evaluating IR methods. Each point represents an algorithm. Higher correlations indicates better performance in evaluating perceptual image algorithms.}
    \label{fig:scatter_plots_alg}
\end{figure*}

\begin{figure*}
    \centering
    \includegraphics[width=\linewidth]{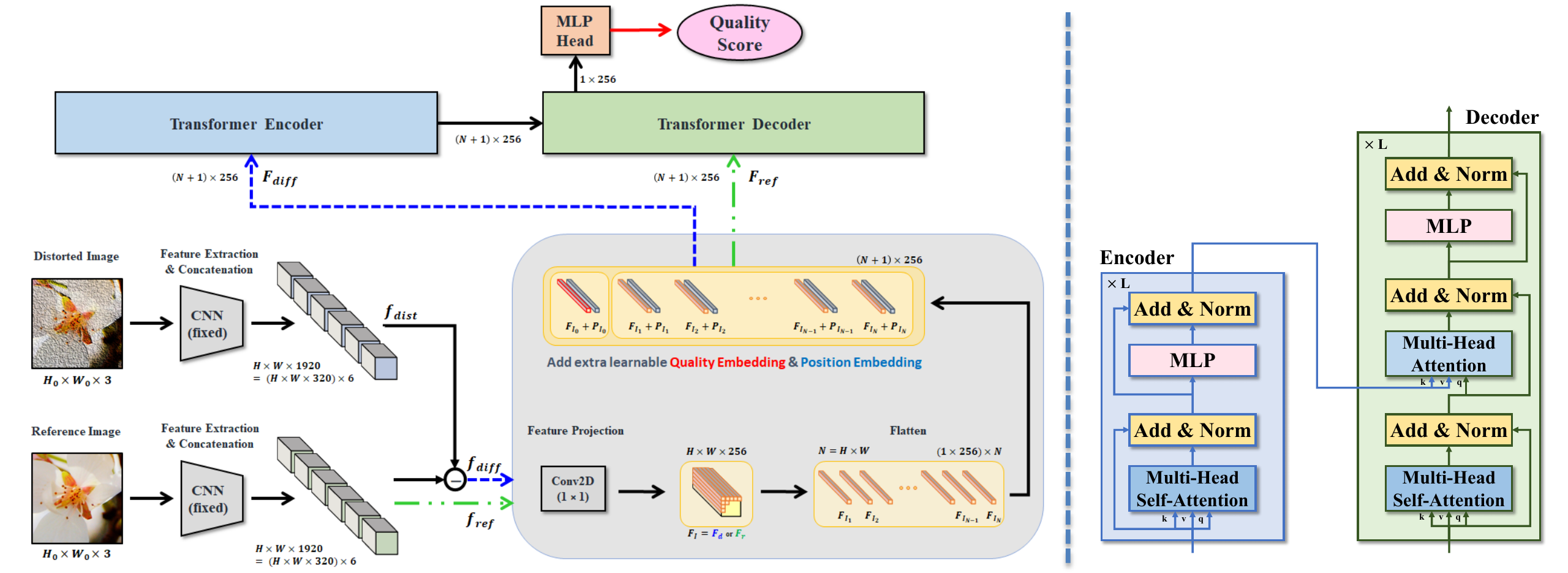}
    \caption{The overview of LIPT team's Image Quality Transformer (IQT) method.}
    \label{fig:5.1}
\end{figure*}

\paragraph{Evaluation protocol.}
Our evaluation indicator, namely main score, consists of both Spearman rank-order correlation coefficient (SRCC) \cite{sheikh2006statistical} and Person linear correlation coefficient (PLCC) \cite{benesty2009pearson}:
\begin{equation}
	\mathrm{Main\ Score}=\mathrm{SRCC}+\mathrm{PLCC}.
\end{equation}
The SRCC evaluates the monotonicity of methods that whether the scores of high-quality images are higher (or lower) than low-quality images.
The PLCC is often used to evaluate the accuracy of methods \cite{sheikh2006statistical,gu2020image}.
Before calculating PLCC index, we perform the third-order polynomial nonlinear regression as suggested in the previous works \cite{tid2013,pipal}.
By combining SRCC and PLCC, our indicator can measure the performance of participating models in an all-round way.

\paragraph{Challenge Phases.}
The whole challenge consists of three phases: the developing phase, the validation phase, and the testing phase.
In the developing phase, the participants can access to the reference and distorted images of the training set and also the MOS labels.
This period is for the participants to familiarize themselves with the structure of the data and develop algorithms.
In the validation phase, the participants can access the reference and distorted images of the training set and no labels are provided.
The participants had the opportunity to test their solutions on the validation images and to receive immediate feedback by uploading their results to the server.
A validation leaderboard is available.
In the testing phase, the participants can access to the reference and distorted images of the training set.
A final predicted perceptual similarity result is required before the challenge deadline.
The participants also need to submit the executable file and a detailed description file of the proposed method. 
The final results were then made available to the participants.


\begin{figure*}
    \centering
    \includegraphics[width=0.9\linewidth]{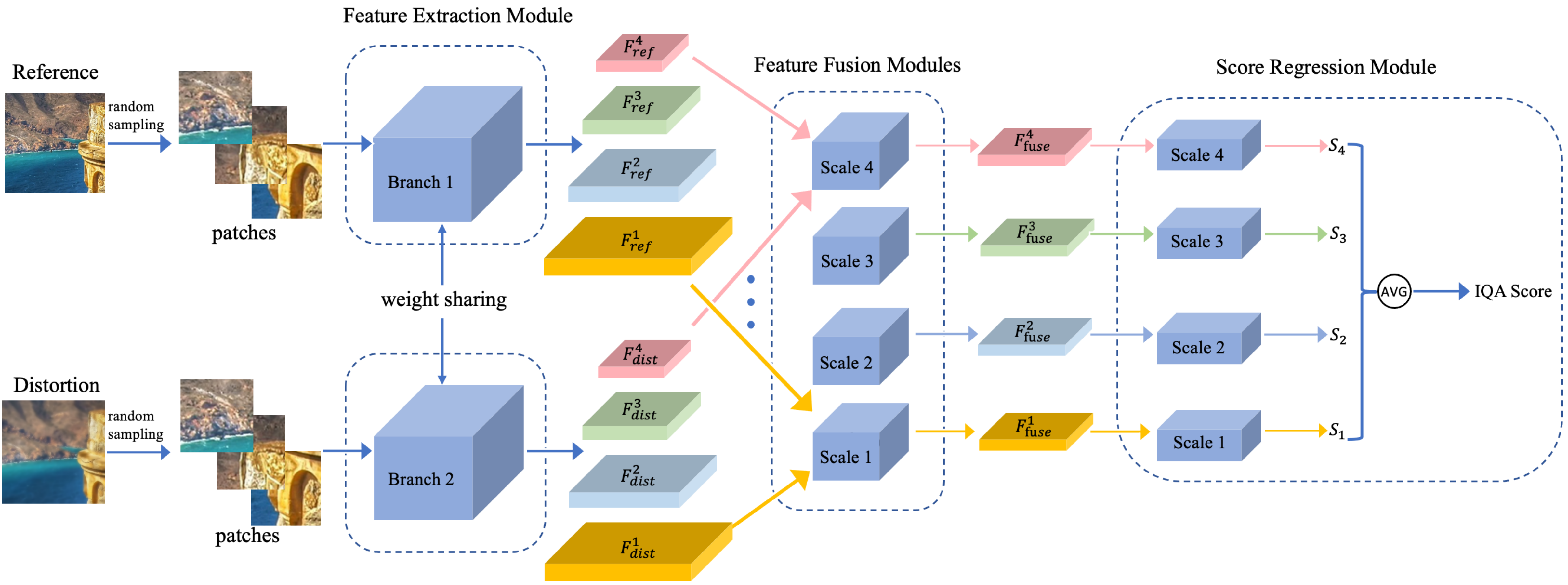}
    \caption{The overview of MT-GTD team's bilateral-branch multi-scale image quality estimation (IQMA) network.}
    \label{fig:5.2}
\end{figure*}

\begin{figure}
    \centering
    \begin{tabular}{c}
        \includegraphics[width=0.95\linewidth]{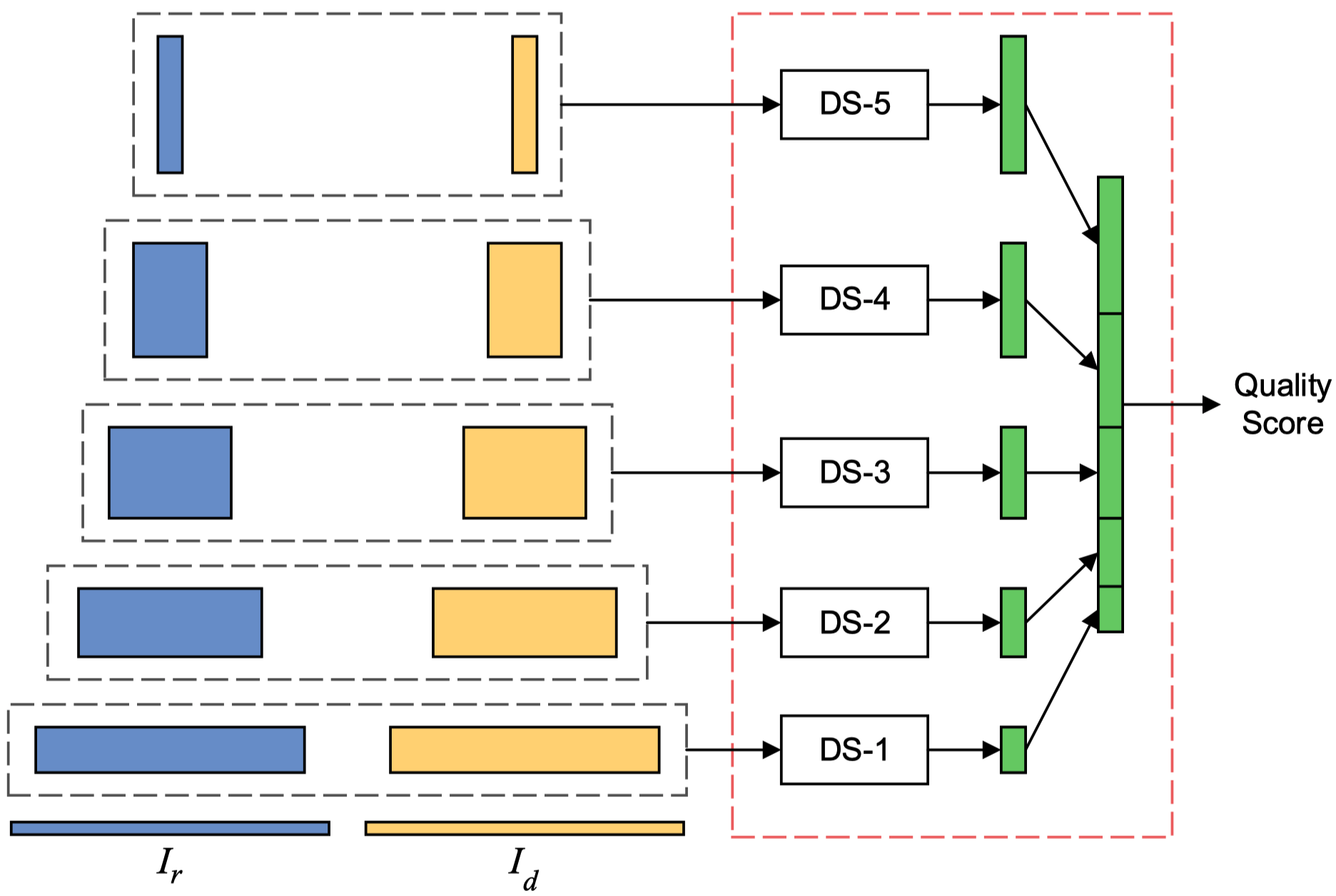}\\
        (a) The overall framework.\\
        \includegraphics[width=0.95\linewidth]{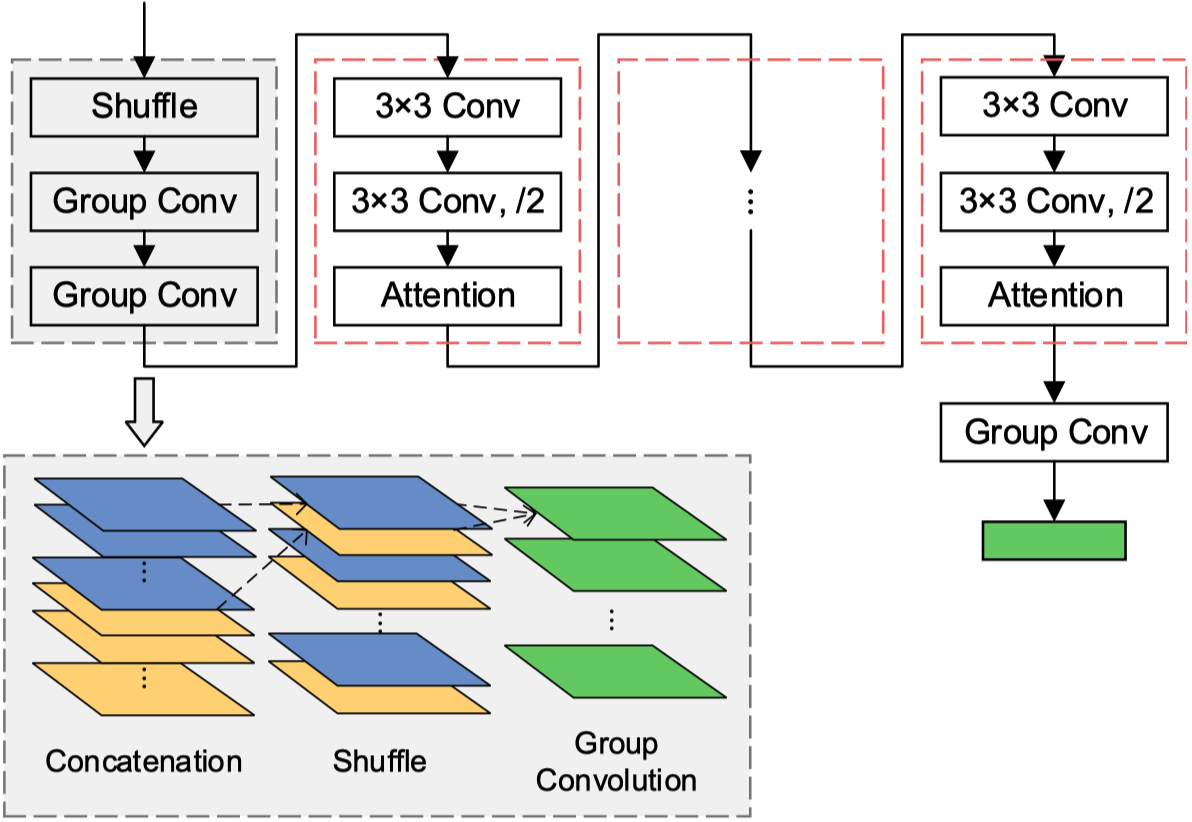}\\
        (b) The proposed deep similarity module.\\
    \end{tabular}
    \caption{The Amaurotia team: Learning to Learn a perceptual image path similarity metric.}
    \label{fig:5.3}
\end{figure}

\begin{figure*}
    \centering
    \includegraphics[width=\linewidth]{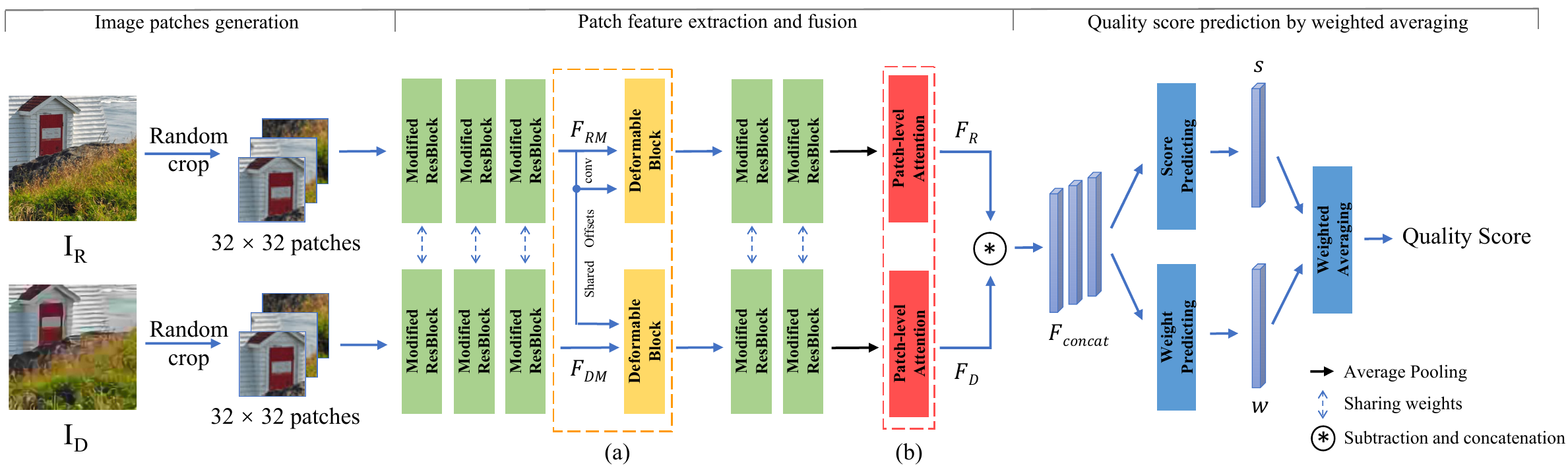}
    \caption{The overview of THUIIGROUP1919 team's Region Adaptive Deformable Network (RADN).}
    \label{fig:5.4}
\end{figure*}

\begin{figure}
    \centering
    \begin{tabular}{c}
        \includegraphics[width=0.95\linewidth]{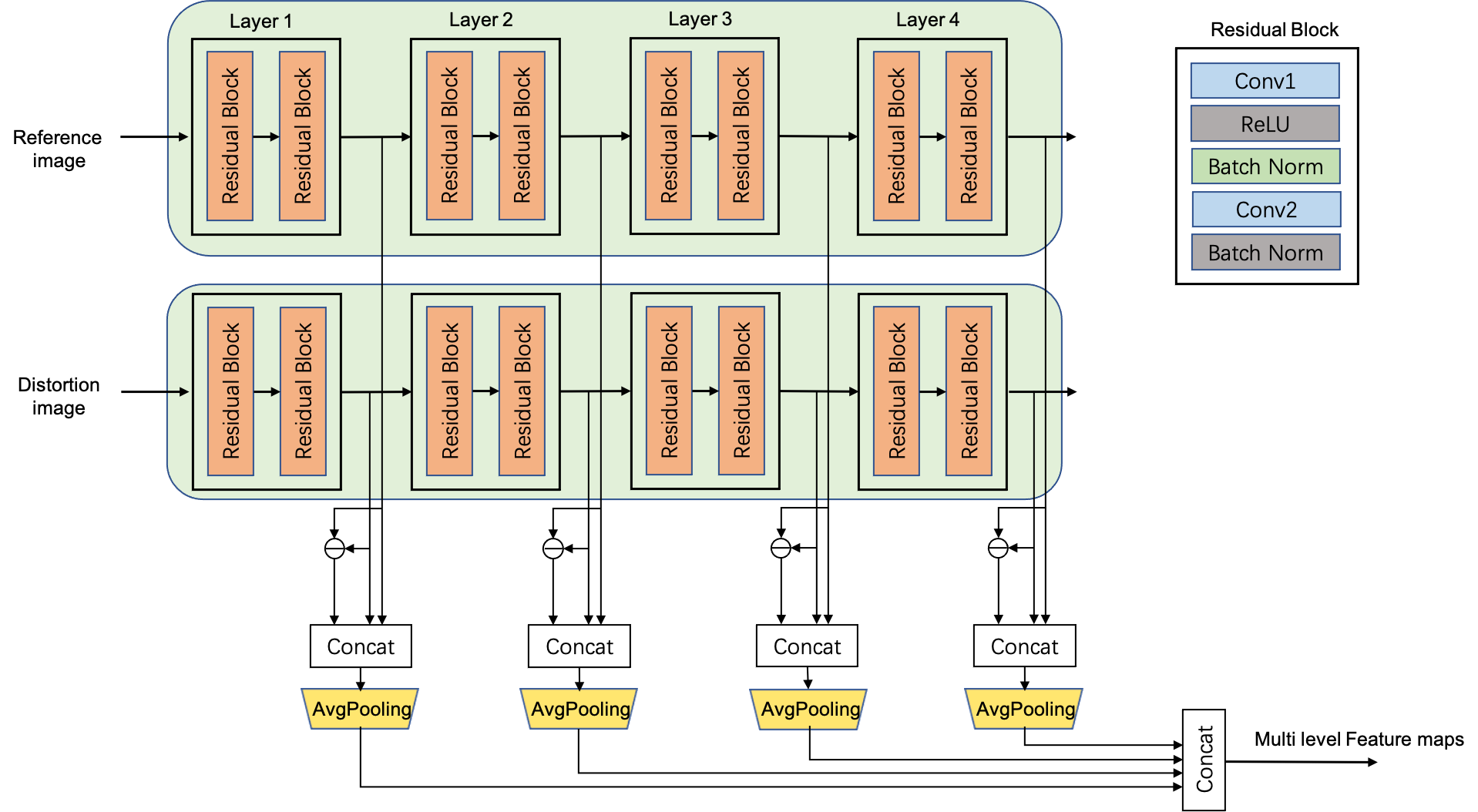}\\
        (a) Framework for feature extraction.\\
        \includegraphics[width=0.95\linewidth]{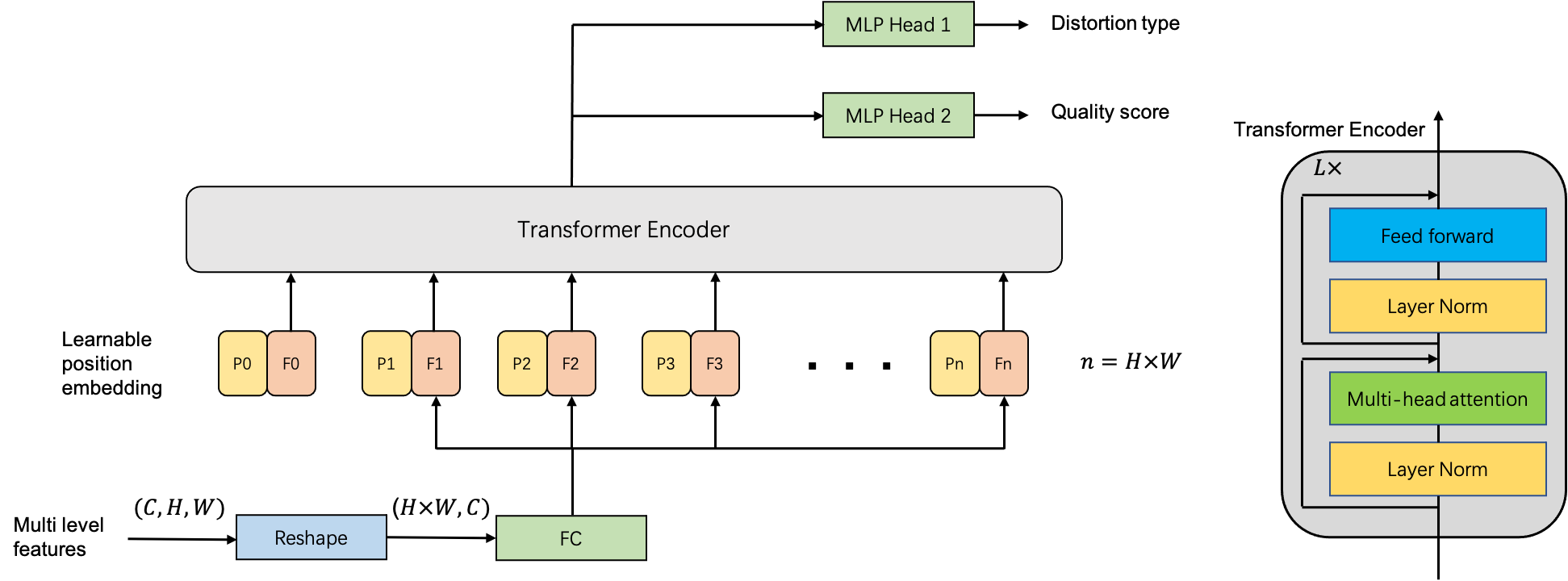}\\
        (b) Framework for transformer.\\
    \end{tabular}
    \caption{The Yahaha team: A Transformer-based perceptual image quality assessment framework leveraging multi level features.}
    \label{fig:5.5}
\end{figure}

\section{Challenge Results}
There are 13 teams participated in the testing phase of the challenge.
\tablename~\ref{tab:main_results} reports the main results and important information of these teams.
We also select some existing representative IQA methods as our baseline.
Specifically, We choose PSNR, SSIM \cite{ssim} and FSIM \cite{fsim} as representative traditional IQA methods, PI \cite{blau2018perception} and NIQE \cite{niqe} as representative blind IQA methods, and LPIPS \cite{zhang2018unreasonable}, DISTS \cite{dists}, PieAPP \cite{prashnani2018pieapp} and SWD \cite{gu2020image} as representative deep-learning based methods.
\tablename~\ref{tab:main_results} reports the final test results and rankings of the challenge.
The methods are briefly described in Section~\ref{sec:methods} and the team members are listed in Appendix~\ref{sec:apd:team}.

As shown in \tablename~\ref{tab:main_results}, 11 of 13 participating teams achieve an SRCC score higher than 0.75 on PIPAL, which significantly surpasses the highest performance of existing algorithms (0.65).
The champion team achieves an SRCC score of 0.799 and a PLCC score of 0.790, refreshing the state-of-the-art performance on PIPAL.
In order to evaluate their performance on traditional distortion types, we also report their results on TID2013 \cite{tid2013} and LIVE \cite{live} dataset in \tablename~\ref{tab:main_results}.
The top three teams in the challenge all achieve competitive results with existing methods on TID2013 and LIVE, showing their good generalization ability on traditional distortion types.

\figurename~\ref{fig:scatter_plots} shows the scatter distributions of subjective MOS scores vs. the predicted scores by the top solutions and the other 5 IQA metrics on PIPAL test set.
The curves shown in \figurename~\ref{fig:scatter_plots} were obtained by a third-order polynomial nonlinear fitting.
One can observe that the objective scores predicted by the top solutions have higher correlations with the subjective evaluations than existing methods.
We then present the analysis of IQA methods as performance measures for perceptual image processing algorithms.
Recall that an important goal of this challenge is to promote more promising IQA metrics for perceptual-oriented algorithms.
In \figurename~\ref{fig:scatter_plots_alg}, we show the scatter plots of subjective scores vs. the top solutions and some commonly-used IQA metrics for some perceptual-oriented algorithms.
As can be seen, the top solutions perform well in evaluating the images in the testing set.
Among them, the correlation between the evaluation of the champion solution (1st) and the subjective score reaches 0.95.

\section{Challenge Methods}
\label{sec:methods}
We describe the submitted solution details in this section.
\subsection{LIPT}
LIPT team is the winner of this challenge.
They develop an image quality transformer (IQT), introduced in \cite{IQT2021ntire}, that applies a transformer architecture to the perceptual IQA task.
Recently, the transformer-based models achieve impressive results in many vision tasks \cite{dosovitskiy2020image,khan2021transformers}.
However, this is the first time that the transformer technique \cite{vaswani2017attention} has been applied to the full-reference IQA task.
The overview framework of their IQT method is illustrated in \figurename~\ref{fig:5.1}.
The IQT method consists of three parts, the feature extraction network, the transformer encoder and decoder, and the prediction head.
Firstly, an Inception-ResNet-V2 network \cite{szegedy2017inception} pre-trained on ImageNet \cite{russakovsky2015imagenet} is used to extract perceptual representations from both reference and distorted images.
The extracted feature maps are then projected to vectors and a trainable extra quality embedding and position embedding are also added.
Secondly, the transformer encoder takes the embedding of the feature map difference calculated between the reference feature map and the distorted feature map as input, and the output of the encoder is sent to the transformer decoder together with the embedded feature map of the reference image.
The transformer encoder and decoder are based on the standard architecture of the transformer, which consists of multi-head attention modules, multi-layer perceptions and layer normalization.
At last, the prediction head takes the output of the transformer decoder as the input and predicts the perceptual similarity score.

In the training phase of the IQT, $M$ overlapping image patches of size $256\times256$ are cropped from both the reference image and the distorted image.
The final quality score is obtained by averaging the quality scores of these patches.
Horizontal flip and random rotation are applied as data augmentation during the training
And the loss is calculated using a mean squared error between the predicted scores and the ground truth scores.

\subsection{MT-GTD}
MT-GTD team wins the second place in our challenge.
They contribute a new bilateral-branch multi-scale image quality estimation (IQMA) network, which is detailed in \cite{IQMA2021ntire}.
At first, ResNet \cite{he2016deep} pre-trained on ImageNet \cite{krizhevsky2012imagenet} is used as the feature extraction backbone.
The IQMA network has two branches with Feature Pyramid Network (FPN)-like architecture to extract multi-scale features from patches of the reference image and corresponding patches of the distorted image separately.
The feature maps of the same scale from both branches are then sent into several scale-speciﬁc feature fusion modules.
Each module performs both a feature fusion operation and a pooling operation for corresponding features.
Then several score regression modules are used to learn a quality score for each scale.
Finally, image scores for different scales are fused as the quality score of the image.
The overall framework is illustrated in \figurename~\ref{fig:5.2}.
It is worth noting that the parallel integration method is used for ensemble models in the challenge.
The MT-GTD team ensembles 9 models which are depicted in the testing description with an average score of all these models as the ﬁnal score.

In the training phase of the IQMA network, a data augmentation operation is specifically designed to address the imbalance issue in the PIPAL training set.
They observe that only a few images in the PIPAL training set have a subjective score of less than 1300.
They re-sample these images using random horizontal flipping, vertical flipping and random rotations as data augmentation.
The augmentation operation successfully increases the percentage of the images that have very small subjective scores.
The training is driven using smooth $L_1$ loss as it is more robust to outliers.

\subsection{The Amaurotia}
The Amaurotia team extends the LPIPS metric \cite{zhang2018unreasonable} and proposed Learning to Learn a Perceptual Image Patch Similarity (L$^2$PIPS) method employed a new deep similarity (DS) module.
\figurename~\ref{fig:5.3} (a) shows the overall framework of the proposed method.
A ResNet-50 \cite{he2016deep} network pre-trained on ImageNet \cite{krizhevsky2012imagenet} is used as the feature extraction backbone and multi-scale deep representations are extracted for comparison.
Their main novelty lies in the DS module, which is designed to predict the similarities between feature pairs.
In each DS module, a channel shuffle operation and group convolution are performed to compare the feature channels pair-wisely, as shown in \figurename~\ref{fig:5.3} (b).
Moreover, frequency channel attention (FCA) \cite{qin2020fcanet} technique is also employed to perform attention mechanism.
In the challenge, an additional cosine similarity is also used as an auxiliary output if each DS module.

The training phase of the proposed L$^2$PIPS has two stages.
In the first 100 epochs, the ResNet-50 feature extraction network is frozen and the loss function is the $L_1$ loss.
In the second phase, another 100 epochs training is performed.
The weights of the feature extraction network are also trainable in this stage and the loss function is changed to mean square error loss.

\begin{figure}
    \centering
    \includegraphics[width=0.8\linewidth]{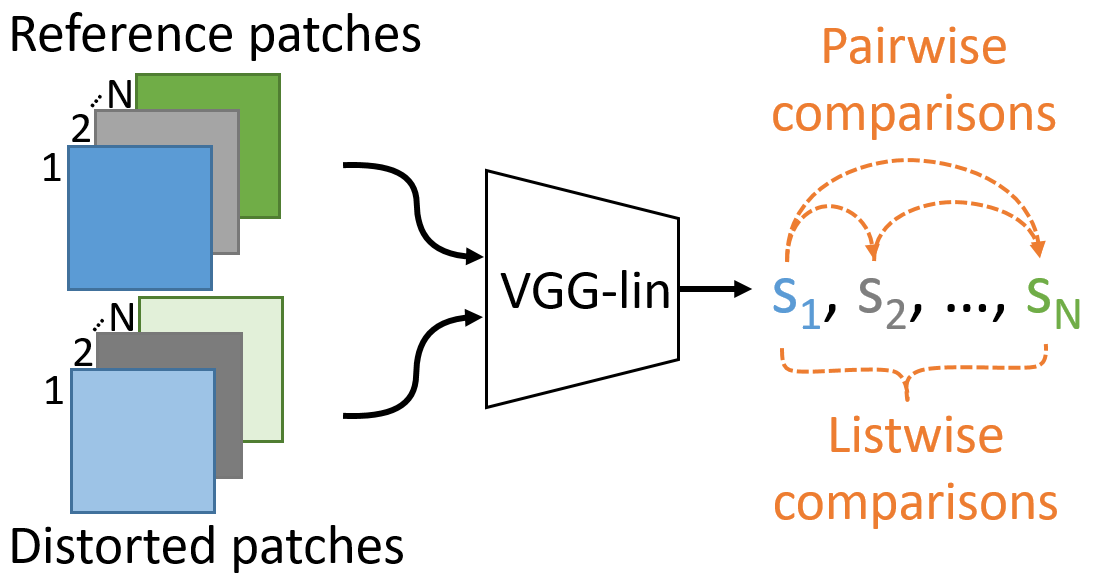}
    \caption{The Huawei Noah’s Ark team: Image agnostic pairwise comparisons and differentiable correlation loss functions for full-reference image quality assessment.}  
    \label{fig:5.6}
\end{figure}

\subsection{THUIIGROUP1919}
THUIIGROUP1919 proposes a new IQA network, namely Region Adaptive Deformable Network (RADN) \cite{RADN2021ntire}, which is illustrated in \figurename~\ref{fig:5.4}.
Firstly, they modified the original residual block by removing batch normalization and adopting $3\times3$ convolution instead of $7\times7$.
These modified residual blocks are trained from scratch to extract features for comparison.
Note that no pre-trained model is used in the proposed RADN, which is different from most of the rest solutions.
Considering that the human visual system's low sensitivity to the error and misalignment of the edges in perceptual distortions \cite{gu2020image}, reference-oriented deformable convolution modules are adopted to make better use of the reference information.
The offset parameters are calculated using the reference image and the deformable convolutions use this offset to process the distorted image.
Before being sent to quality score prediction, a novel patch-level attention module is designed to explore the correlations between local patches.

In the training phase of the RADN network, a contrastive pre-training strategy is proposed to make the model learn how to distinguish the image quality rather than directly guess the quality score.
Similar to LPIPS \cite{zhang2018unreasonable} and SWD \cite{gu2020image}, a two-layer fully-connected network is employed to predict pairwise probabilities from the quality scores.
The cross entropy loss function is used to calculate loss between the predicted probabilities and the ground truth probabilities obtained using Elo system in PIPAL dataset.

\subsection{Yahaha}
Yahaha team also adopts the transformer technique \cite{vaswani2017attention} to build their IQA method.
A ResNet-18 \cite{he2016deep} backbone network is first employed for feature extraction.
As shown in \figurename~\ref{fig:5.5} (a), both reference image and distortion image are fed to the backbone network to obtain the feature maps of each layer.
The feature maps and the difference of quality maps are concatenated together as $c_i=r_i\oplus d_i\oplus(r_i-d_i)$, where $c_i$ is the concatenated feature from layer $i$, $r_i$ and $d_i$ are the reference and distorted image feature maps of layer $i$.
Then, feature maps from different layers are downsampled to the same spatial resolution by average pooling before being sent to the transformer.
Structure of transformer stage is shown in \figurename~\ref{fig:5.5} (b), features of each position are added with a learnable position embedding before fed to a standard transformer encoder.
The transformer encoder is connected with two multi-layer fully-connected network heads for the prediction of distortion type and opinion score, respectively.
The loss from these two parts are weighted summed during training.

In the training phase, the Yahaha team employs both $L_1$ loss and a relative-distance loss
In relative-distance loss, the difference between prediction scores and the ground truth subjective scores are compared for each image pair.
It can be formulated by:
\begin{equation*}
    Loss = \sum_{i=1}^N\sum_{j=i+1}^N|(y_i-y_j)-(x_i-x_j)|,
\end{equation*}
where $y_i$ and $x_i$ represent the $i$th ground truth score and prediction score, respectively.

\begin{figure}
    \centering
    \begin{tabular}{c}
        \includegraphics[width=0.95\linewidth]{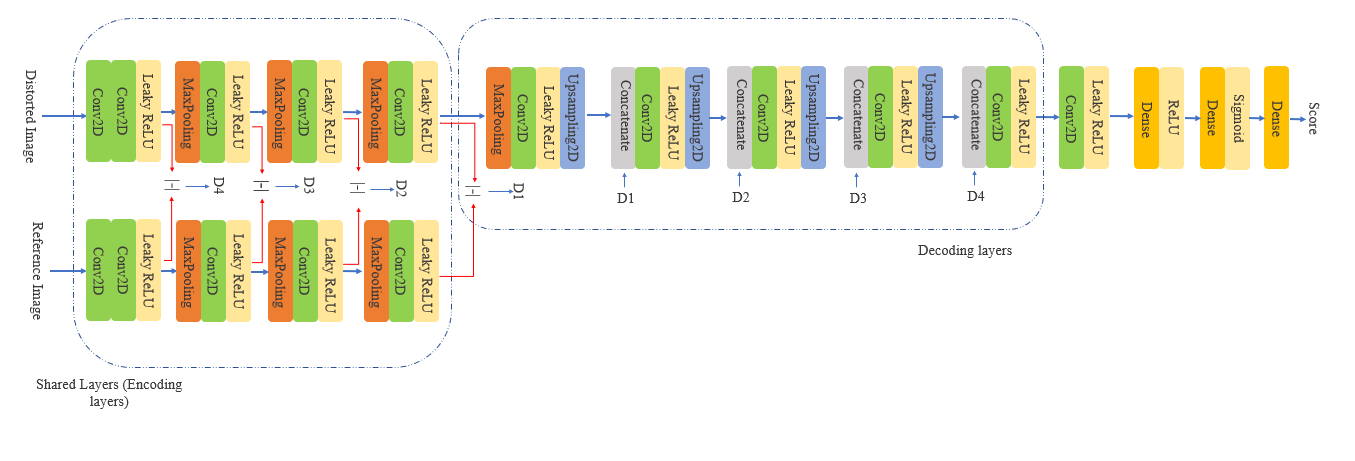}\\
        (a) Siamese-Difference architecture.\\
        \includegraphics[width=0.95\linewidth]{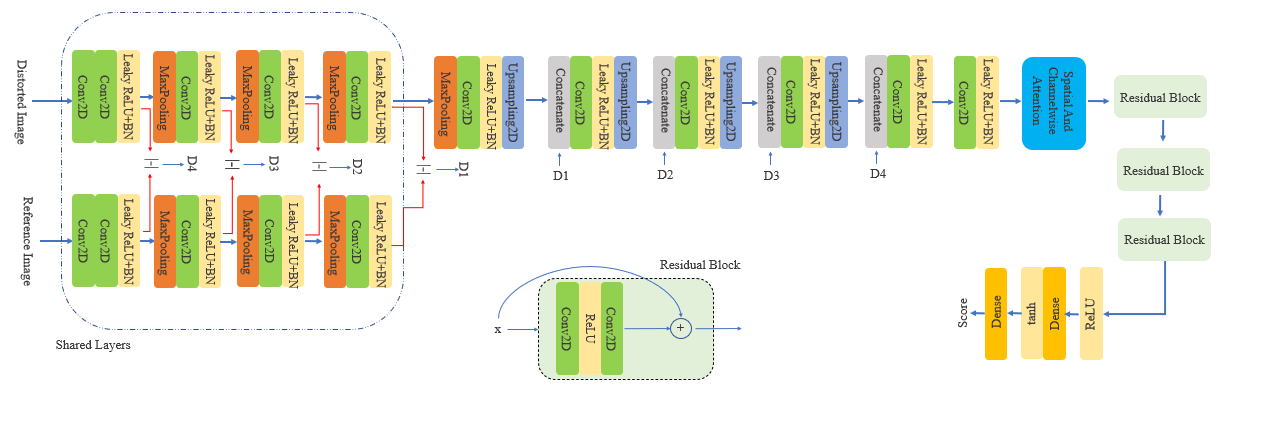}\\
        (b) Siamese-Difference with attention architecture.\\
        \includegraphics[width=0.95\linewidth]{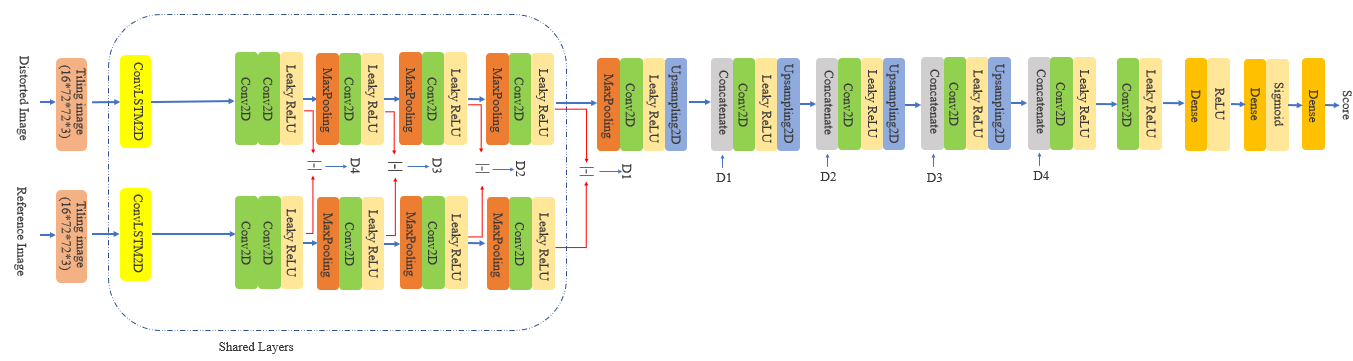}\\
        (c) Siamese-Difference with ConvLSTM layer.\\
        \includegraphics[width=0.95\linewidth]{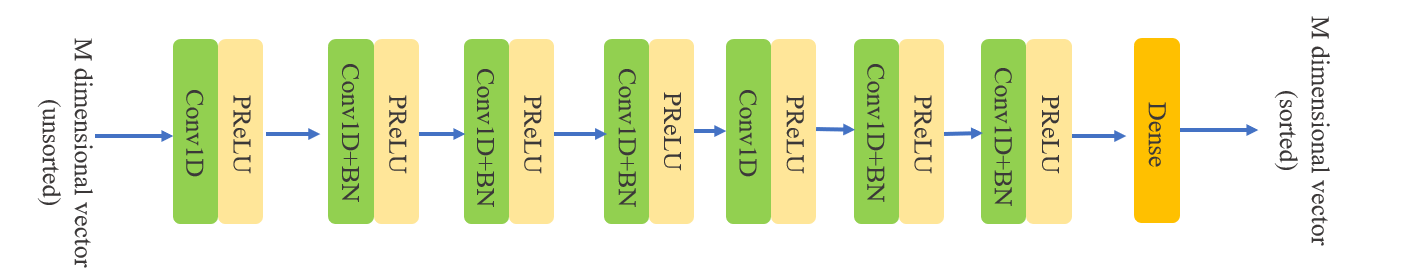}\\
        (c) Ranking model architecture.\\
    \end{tabular}
    \caption{The MACS team: Siamese-Diﬀerence Network for IQA.}
    \label{fig:5.8}
\end{figure}

\begin{figure*}
    \centering
    \includegraphics[width=\linewidth]{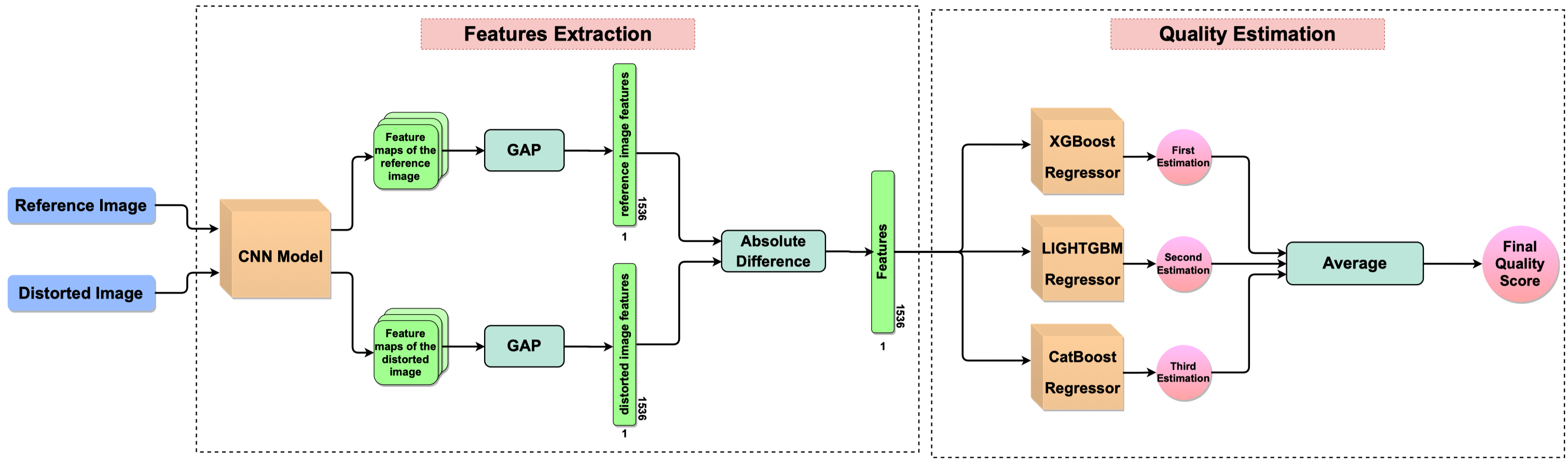}
    \caption{The LION team: Image Quality Estimation based on Ensemble of Gradient Boosting.}
    \label{fig:5.9}
\end{figure*}

\begin{figure}
    \centering
    \includegraphics[width=\linewidth]{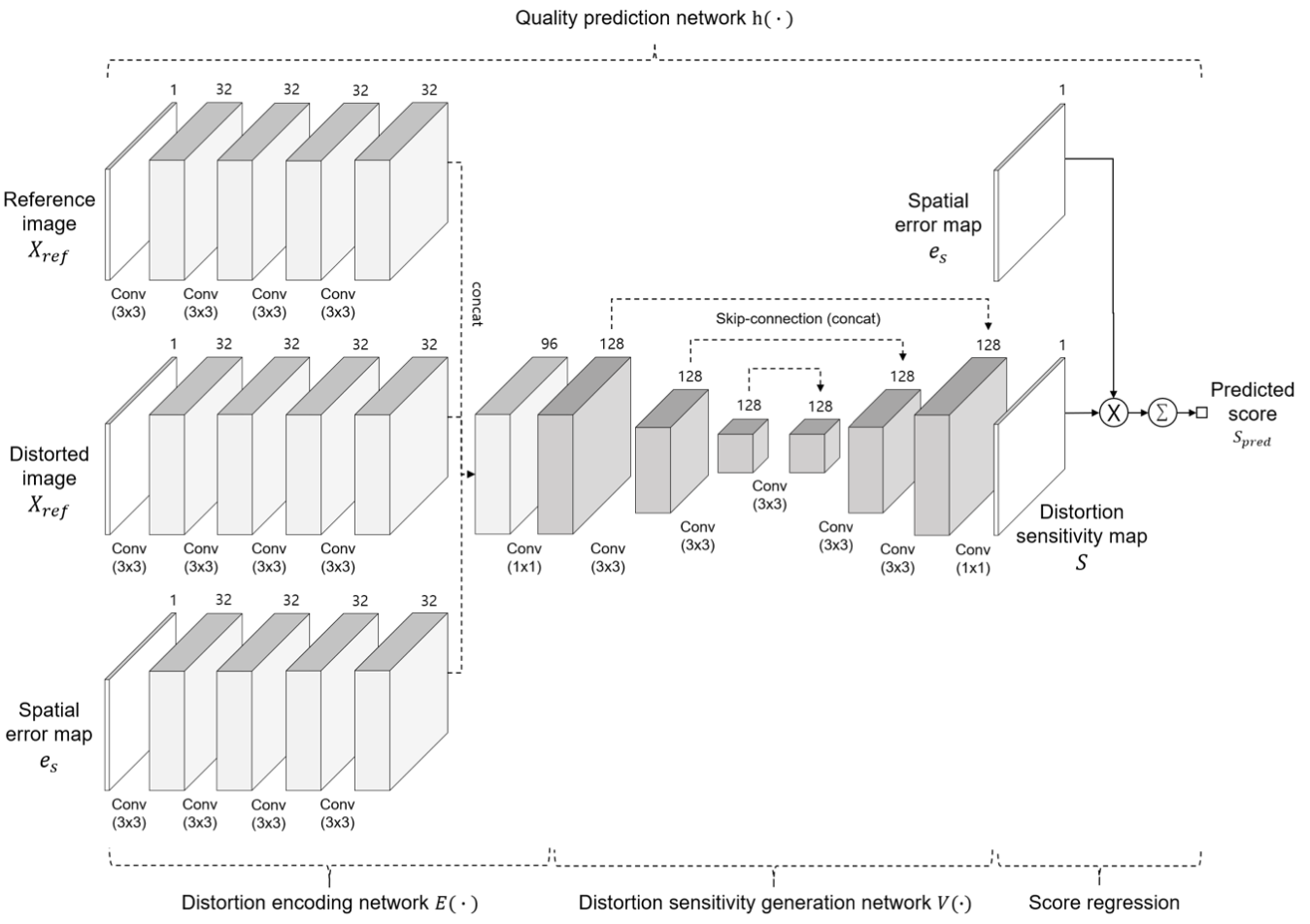}
    \caption{SI Analytics team: Deep Learning-based Distortion Sensitivity Prediction for Full-Reference Image Quality Assessment.}
    \label{fig:5.10}
\end{figure}

\subsection{Huawei Noah's Ark}
The innovative point of the method proposed by Huawei Noah's Ark team is the use of two special loss functions.
The model is trained with a loss function that incorporates (a) pairwise comparisons, and (b) listwise comparisons.
Having many and diverse comparisons with distorted images coming from different reference images provides a better signal during training, and better mimics how IQA is evaluated.

For the pairwise comparison loss, the Bradley-Terry (BT) sigmoid \cite{bradley1952rank} is used to obtain a probability of which image in the pair has the best quality.
The pair of distorted images can originate from different reference images, making the pairwise comparison agnostic to their reference image.
Hence, the model is forced to compare pairs of images with different scene content, as is the case in IQA evaluation.
For training, the MSE loss between the predicted probability from the BT model and its ground truth probability computed with the Elo logistic curve is used.

In addition to the pairwise comparison, novel differentiable regularizers derived from the PLCC and SRCC are also used as losses.
The regularizers act as listwise comparisons where all samples in a mini-batch form an ordered list.
Having listwise comparisons complement the pairwise comparisons, as every sample is now also compared to the rest of the mini-batch.
Hence, the model is forced to learn subtle differences (\ie, pairwise comparisons) as well as relative differences (\ie, listwise comparisons).
These two losses have the same weights in the training phase.

\subsection{debut\_kele}
Debut\_kele team employs ensemble technology to build their method.
They mainly employs the complementary information provided by traditional IQA and deep learning-based IQA algorithms.
The contributed solution can be divided into two main parts: feature extraction using different perceptual image quality assessment, which include SSIM, MS-SSIM, CW-SSIM, GMSD, LPIPS-VGG, DISTS, NLPD, FSIM, VSI, VIFs, VIF, and MAD.
Then, a regression model is built using XGBoost \cite{chen2016xgboost} based on the pre-calculated IQA.
The 'max depth' parameter of the model is set to 3 and learning rate was set at 0.01.
In addition, feature subsample and the sample subsample values were set at 0.7 to prevent from overﬁtting.
The maximum iteration round is set as 10000 while the early-stopping round is set 500.
In the challenge, 5 fold bagging-based ensemble strategy is used.

\subsection{MACS}
MACS team proposes three different networks that all based on a Siamese-Difference architecture, which is detailed in \cite{asna2021ntire}.
The basic Siamese-Difference network are illustrated in \figurename~\ref{fig:5.8} (a).
In the decoding part, the absolute values of the outputs are concatenated from the encoding part.
Instead of using concatenation of the encoded images directly for the skip connections, the absolute difference of the encoded feature maps have been used.
The decoding part is not symmetric, which is different from the traditional Siamese networks
The authors argue that this architecture has more capability in representing useful features to estimate the quality score rather than the traditional Siamese network architecture.
In the second network, channel-wise attention and spatial attention are used to improve the Siamese-Diﬀerence architecture, as shown in \figurename~\ref{fig:5.8} (b).
To mitigate the issue of high computational cost of the proposed network on big images, the third network first tiled each input into patches with size 72 × 72 and stack them as an image bag.
These patches are sent to a ConvLSTM Layer, as shown in \figurename~\ref{fig:5.8} (c).
In the challenge, each network is trained separately, and the averaged score of them are used as the final score.

In the training phase, multiple loss functions are combined to achieve better results.
Firstly, MSE loss is used as the most common loss function for a regression task.
A new differenitable pearson loss is also employed to increase the PLCC performance.
At last, a novel rank loss is proposed in order to improve the SRCC rank correlation performance.
Inspired by SoDeep \cite{engilberge2019sodeep}, a ranker network is trained to learn how to sort $M$ inputs.
The network architecture is shown in \figurename~\ref{fig:5.8} (d).
To train this network, $M$ dimensional random vectors are generated and the network is trained to learn how to sort the inputs by minimizing mean absolute error between the sorted input and the output of the network.
After training this network, this network can be used as a differentiable sorting function.
Suppose $R(S)$ and $R(S')$ are the estimated ranked vectors of $S$ and $S'$, respectively.
The squared difference between these two vectors is used in order to increase SRCC measure as $L_{Rank}=\frac{1}{M}\|R(S)-R(S')\|^2$.

\subsection{LION}
LION team also employs boosting strategy to build their solution, which is introduced in \cite{EGB2021ntire}.
Firstly, the VGG-16 network \cite{vgg} pre-trained on the ImageNet \cite{krizhevsky2012imagenet} is used as the feature extraction module.
First, an analysis is performed on features extracted from different layers to determine if a particular layer of the model provides more relevant feature maps for image quality assessment task.
For each layer of the VGG-16 model, a distance between reference and distorted images is computed using extracted feature maps.
The correlation between the calculated distances and the subjective scores is computed using SRCC.
Three intermediate layers (block4 conv2, block4 conv3 and block5 conv1) are highlighted by analyzing these SRCC values and are considered to be the best candidates for the evaluation of the perceptually image quality.
When performing image quality estimation, the reference and distorted images are fed to the VGG-16 network to derive the feature vector from the three selected intermediate convolution layers.
Then, global average pooling is performed to reduce the feature maps into a manageable size.
Finally, to provide the feature vector to the regression part of the framework, the absolute difference between the extracted features of the reference and distorted images is calculated.
The feature vector is then fed into the three Gradient Boosting regression models, which are XGBoost \cite{chen2016xgboost}, LightGBM \cite{ke2017lightgbm} and CatBoost \cite{dorogush2018catboost}, to be regressed to predict three image quality scores. The three models are ensemble models using decision trees.
The idea behind using three Gradient Boosting regression models, is to combine the predictions of these models to form an ensemble that outperforms a single trained model. 

\subsection{SI Analytics}
SI Analytics team proposes a method motivated by the visual sensitivity map, which is detailed in \cite{DeepQA2021ntire}. 
The visual sensitivity map refers to a weighting map describing the degree of visual importance of each pixel to the human visual system (HVS).
As shown in \figurename~\ref{fig:5.10}, their method firstly predicts visual sensitivity map allocating local weights to the pixels according to their local spatial characteristics based on the reference image, the distorted image, and the spatial error map.
The spatial error map is defined by 
$
\frac{    log(    1/  ((X_{ref}-X_{dis} )^{2} + \varepsilon/ 255^{2}) )  }{   log(255^{2}/\varepsilon) }
$,
where $X_{ref}$ and $X_{dis}$ are gray-scale reference image and distorted image normalized to $[0,1]$, and $\varepsilon = 1$.
With the spatial error map and predicted visual sensitivity map, the Hadamard product and global average pooling are applied in series to obtain the final subjective score.
Different from DeepQA \cite{kim2017deep}, they use UNet \cite{unet} structure to conserve the spatial information of input images instead of using down-sampling operations for predicting distortion sensitivity maps.
Moreover, there are more convolution layers in the encoding network to enlarge the receptive fields.

In the training phase, they normalize the ground-truth score into $[0,1]$ and oversample those images outside the range of $[0.4, 0.8]$ to relieve score imbalance problems.
The loss function adopts the combination of mean-square error and L2 loss between predicted quality score and subjective scores.

\subsection{tsubota}
Tsubota team builds their method by modifying PieAPP \cite{prashnani2018pieapp} on feature extraction network and pooling layers. 
Specifically, they replace the VGG \cite{vgg} network in PieAPP with AlexNet \cite{krizhevsky2012imagenet} pre-trained on ImageNet \cite{deng2009imagenet} image and adopt Blurpool \cite{blurpool} to avoid aliasing, not commonly used Maxpool. 

In the training phase, they firstly train their model on the PieAPP dataset to obtain a pre-trained model. 
Then, they finetune the pre-trained model on the PIPAL dataset.
For those trainings, the input image was randomly cropped into 36 patches with a resolution of 64 $\times$ 64 pixels. 
In the testing phase, they use a 64 $\times$ 64 sliding window with the stride size set to 6 pixels.
The rest of settings are same with PieAPP.

\section*{Acknowledgements}
We thank the NTIRE 2021 sponsors: Huawei,
Facebook Reality Labs, Wright Brothers Institute, MediaTek, OPPO and ETH Zurich (Computer Vision Lab).

\appendix
\section{Teams and Affiliations}
\label{sec:apd:team}

\subsection*{NTIRE 2021 Team}
\noindent\textit{\textbf{Title: }}\\ NTIRE 2021 Challenge on Perceptual Image Quality Assessment\\
\noindent\textit{\textbf{Members:}}\\ \textit{Jinjin Gu$^1$ (jinjin.gu@sydney.edu.au)}, Haoming Cai$^2$, Chao Dong$^2$,  Jimmy S. Ren$^3$, Yu  Qiao$^2$, Shuhang Gu$^1$, Radu Timofte$^4$\\
\noindent\textit{\textbf{Affiliations: }}\\
$^1$ School of Electrical and Information Engineering, The University of Sydney\\
$^2$ Shenzhen Institutes of Advanced Technology, Chinese Academy of Sciences\\
$^3$ SenseTime Research\\
$^4$ Computer Vision Lab, ETH Zurich, Switzerland\\

\subsection*{LIPT}
\noindent\textit{\textbf{Title:}}\\ Perceptual Image Quality Assessment with Transformers\\
\noindent\textit{\textbf{Members: }}\\ \textit{Manri Cheon$^1$ (manri.cheon@lge.com)}, Sungjun Yoon$^1$, Byungyeon Kang$^1$, Junwoo Lee$^1$\\
\noindent\textit{\textbf{Affiliations: }}\\
$^1$ LG Electronics\\

\subsection*{MT-GTD}
\noindent\textit{\textbf{Title:}}\\IQMA Network: Image Quality Multi-Scale Assessment Network\\
\noindent\textit{\textbf{Members:}}\\ \textit{Qing Zhang$^1$ (zhangqing31@meituan.com)}, Haiyang Guo$^1$, Yi Bin$^1$, Yuqing Hou$^1$, Hengliang Luo$^1$\\
\noindent\textit{\textbf{Affiliations: }}\\
$^1$ Meituan Group\\

\subsection*{The Amaurotia}
\noindent\textit{\textbf{Title: }}\\ 
L$^2$PIPS: Learn to Learn a Perceptual Image Patch Similarity Metric\\
\noindent\textit{\textbf{Members:}} \\ 
\textit{Jingyu Guo$^1$ (gjy19@mails.tsinghua.edu.cn)}, Zirui Wang$^1$, Hai Wang$^1$, Wenming Yang$^2$\\
\noindent\textit{\textbf{Affiliations: }}\\
$^1$ Department of Electronic Engineering, Tsinghua University\\
$^2$ Shenzhen International Graduate School, Department of Electronic Engineering, Tsinghua University\\

\subsection*{THUIIGROUP1919}
\noindent\textit{\textbf{Title: }} \\
WResNet: A Light Residual Network for FR-IQA\\
\noindent\textit{\textbf{Members:}} \\
\textit{Qingyan Bai$^1$ (baiqingyan1998@163.com), Shuwei Shi$^1$ (ssw20@mails.tsinghua.edu.cn)}, Weihao Xia$^1$, Mingdeng Cao$^2$, Jiahao Wang$^2$, Yifan Chen$^1$, Yujiu Yang$^1$\\
\noindent\textit{\textbf{Affiliations: }}\\
$^1$ Tsinghua Shenzhen International Graduate School, Tsinghua University\\
$^2$ Department of Automation, Tsinghua University

\subsection*{Yahaha}
\noindent\textit{\textbf{Title: }} \\
A Transformer-based perceptual image quality assessment framework leveraging multi level features\\
\noindent\textit{\textbf{Members:}} \\ 
\textit{Yang Li$^1$ (liyang.00@pku.edu.cn)}, Tao Zhang$^2$, Longtao Feng$^2$, Yiting Liao$^2$, Junlin Li$^2$\\
\noindent\textit{\textbf{Affiliations: }}\\
$^1$ Peking University\\
$^2$ Bytedance Inc.

\subsection*{Huawei Noah's Ark}
\noindent\textit{\textbf{Members:}} \\ 
\textit{William Thong$^1$ (william.thong@huawei.com)}, Jose Costa Pereira$^1$, Ales Leonardis$^1$, Steven McDonagh$^1$\\
\noindent\textit{\textbf{Affiliations: }}\\
$^1$ Huawei Noah’s Ark Lab

\subsection*{debut\_kele}
\noindent\textit{\textbf{Members:}} \\ 
\textit{Kele Xu$^1$ (kelele.xu@gmail.com)}, Lehan Yang$^2$, Hengxing Cai$^3$, Pengfei Sun$^{45}$\\
\noindent\textit{\textbf{Affiliations: }}\\
$^1$ Key Laboratory for Parallel and Distributed Processing\\
$^2$ The University of Sydney\\
$^3$ 4Paradigm Inc.\\
$^4$ University of Zurich\\
$^5$ ETH Zurich

\subsection*{MACS}
\noindent\textit{\textbf{Title:}}\\IQA Using SIDIS: Siamese Diﬀerential Network with Surrogate Loss function\\
\noindent\textit{\textbf{Members:}}\\ \textit{Seyed Mehdi Ayyoubzadeh$^1$ (ayyoubzs@mcmaster.ca)}, Ali Royat$^2$\\
\noindent\textit{\textbf{Affiliations: }}\\
$^1$ ECE Department, McMaster University\\
$^2$ EE Department, Sharif University of Technology\\

\subsection*{LION}
\noindent\textit{\textbf{Title:}}\\
Image Quality Estimation based on Ensemble of Gradient Boosting\\
\noindent\textit{\textbf{Members:}}\\ \textit{Sid Ahmed Fezza$^1$ (sfezza@inttic.dz)}, Dounia Hammou$^1$, Wassim Hamidouche$^2$\\
\noindent\textit{\textbf{Affiliations: }}\\
$^1$ National Institute of Telecommunications and ICT, Oran, Algeria\\
$^2$ Univ. Rennes, INSA Rennes, CNRS, IETR - UMR 6164, Rennes, France\\

\subsection*{SI Analytics}
\noindent\textit{\textbf{Title:}}\\
Deep Learning-based Distortion Sensitivity Generation for Full-Reference Image Quality\\
\noindent\textit{\textbf{Members:}}\\ \textit{Sewoong Ahn$^1$ (sfezza@inttic.dz)}, Kwangjin Yoon$^1$\\
\noindent\textit{\textbf{Affiliations: }}\\
$^1$ SI Analytics

\subsection*{tsubota}
\noindent\textit{\textbf{Title:}}\\
PieAPP Using ImageNet Pre-trained AlexNet\\
\noindent\textit{\textbf{Members:}}\\ \textit{Koki Tsubota$^1$ (tsubota@hal.t.u-tokyo.ac.jp)}, Hiroaki Akutsu$^2$, Kiyoharu Aizawa$^1$\\
\noindent\textit{\textbf{Affiliations: }}\\
$^1$ The University of Tokyo\\
$^2$ Hitachi, Ltd.

\section{Change Log}
\label{sec:apd:change}
\paragraph{v1}
Initial pre-print release.
\paragraph{v2}
Correct the incorrect member name in the LIPT team.
The author ``Byungyeon Kangg Kang'' should be ``Byungyeon Kang''.
We update the authorship and the teams information in Sec~\ref{sec:apd:team}.
\paragraph{v3}
Update the figure of the framework in the LIPT team to consonant with their latest version.
Correct the incorrect member name in the SI Analytics team.
The author ``Gwangjin Yoon'' should be ``Kwangjin Yoon''.
Correct the incorrect contact email in the MACS team.
The email ``zhangqing31@meituan.com'' should be ``ayyoubzs@mcmaster.ca''.
We update the authorship and the teams information in Sec~\ref{sec:apd:team}.

{\small
\bibliographystyle{ieee_fullname}
\bibliography{egbib}
}

\end{document}